\documentclass[conference]{IEEEtran}
\IEEEoverridecommandlockouts
\usepackage{cite}
\usepackage{amsmath,amssymb,amsfonts}
\usepackage{algorithmic}
\usepackage{graphicx}
\usepackage{textcomp}
\usepackage{xcolor}
\usepackage{subfig}
\usepackage{multirow}

\def\BibTeX{{\rm B\kern-.05em{\sc i\kern-.025em b}\kern-.08em
    T\kern-.1667em\lower.7ex\hbox{E}\kern-.125emX}}
\begin{document}
\title{Predicting Software Reliability in Softwarized Networks
}

\author{\IEEEauthorblockN{Hasan Yagiz {\"O}zkan, Madeleine Kaufmann, Wolfgang Kellerer, Carmen Mas-Machuca }
\IEEEauthorblockA{\textit{Chair of Communication Networks} \\
\textit{Technical University of Munich}\\
Munich, Germany \\
yagiz.oezkan@tum.de, madeleine.kaufmann@tum.de, wolfgang.kellerer@tum.de, cmas@tum.de}
}

\maketitle

\begin{abstract}

Providing high quality software and evaluating the software reliability in softwarized networks are crucial for vendors and customers. These networks rely on open source code, which are sensitive to contain high number of bugs. Both, the knowledge about the code of previous releases as well as the bug history of the particular project can be used to evaluate the software reliability of a new software release based on SRGM. 
In this work a framework to predict the number of the bugs of a new release, as well as other reliability parameters, is proposed. An exemplary implementation of this framework to two particular open source projects, is described in detail. The difference between the prediction accuracy of the two projects is presented. Different alternatives to increase the prediction accuracy are proposed and compared in this paper.

\end{abstract}


\section{Introduction}

Delivering high quality software is of utmost importance for software vendors and platforms. For that purpose, they aim at reducing both, the number of undetected and unsolved bugs, to increase the software quality and dependability\cite{visagan2019building}. The software development process is a cyclic process that delivers periodically releases with new functionalities, solved bugs, etc. Each new release potentially has residual bugs (i.e., undetected and unsolved bugs from previous releases as well as from the added code). A software bug can be defined as an anomaly that spoils the normal operation of a software. Hence, bugs are big thread to the reliability of the software.


Furthermore, software development is nowadays being open source. Open source software development strategy offers many advantages in cost, service and support. As a result of these advantages the number of open-source projects and  the number of contributors increase consistently. Big open source projects are mostly based on the contributions from different programmers and/or teams, which develop different parts of the same software. Although it is an efficient and low-cost development process, it is prone to contain higher number of bugs than in closed software development processes\cite{falcao2020relating}.

The problem for both, the developers/vendors of the new release as well as the clients considering its adoption, is how to evaluate the quality and dependability of new releases. For this purpose,  we propose in this paper a methodology to estimate the number of residual bugs of a new software release.  

Most of the open source projects,
adopt bug-tracking systems to efficiently manage any detected software bug \cite{yang2017analyzing}. These bug-tracking systems facilitate the bug detection and resolution by offering transparency on providing information about all the detected bugs, which can be processed by data-mining techniques \cite{vizarreta2020dason}. These techniques can be helpful to create the bug history as well as understand the relation between the bug history and code changes of the different releases.

Software code metrics can be defined as the information about the properties of a source code like for example the complexity of the functions or size of the code. These code metrics have been used to predict quality, security vulnerability, and maintainability of a software \cite{mamun2017correlations}. This information can be easily gathered with help of online tools or programming libraries. Moreover, the history of the code metrics, which relates to the project releases can be calculated and considered in the residual bug estimation as proposed in this work. 

In this paper we present the following contributions: 
\begin{itemize}
    \item We propose a framework to predict the number of bugs in a new software release. This framework combines the bug history and the code metrics history of previous release(s) of a software project.
    \item As some new projects do not have enough previous releases to make an accurate prediction. For this purpose we evaluate the usability of cross-project prediction and evaluate the impact of the prediction accuracy.
    \item As the software development of some projects lasts several years (e.g., the first release of ONAP project was released in 2017 and new releases are still coming), aspects such as trends in the software development strategy or changes in the developer teams can impact significantly the bug estimation. Because of these changes, the usage of all previous releases can be unfavorable for the prediction. The impact of the number of previous releases on the prediction quality will be also investigated. 
\end{itemize} 

After finding the best number of releases to consider and favourable metrics to feed the model, this framework can be used to predict the number of bugs in a new software release before its code freeze, which indicates the beginning of debugging process. 

This paper is structured as follows: Section~\ref{sec:Sota} summarizes the previous works that investigated the relation between bugs and code metrics and/or the effects of bugs on software reliability. In Section~\ref{sec:fra} the framework to predict the number of bugs is discussed. An exemplary implementation of the proposed framework is explained in Section~\ref{sec:imp}. The numerical results of this exemplary implementation are presented in Section~\ref{sec:res}. Finally, the paper concludes with some future work ideas in Section~\ref{sec:con}.

\section{Related Work}
\label{sec:Sota}

In this section works that combines the bug manifestation process and the code metrics are discussed by explaining the used methodologies and their outcomes. This section also introduces works that discuss the relation between bugs and software dependability.

In \cite{rajendren2021predicting} the authors used different machine learning methods like linear regression, decision tree and multilayer perception to estimate the number of bugs in a project. The machine learning models are fed by the  code metrics like number of lines of code (LoC), number of commits. Based on data from 15 different projects from \cite{toth2016public} the authors found that the linear regression model outperforms the other machine learning models and machine learning models can be successfully used to predict the number of bugs before the release of the software. As machine learning model they found that the linear regression model performs better than compared models and machine learning models can be successfully used to predict the number of bugs before the release of the software. Their model only consider the prediction after the release date. A prediction before release date might offer a possibility to the developers to provide more reliable software.

The authors of \cite{gupta2015developing} predicted the number of bugs in a file with code metrics. They compared efficiency of four different code metrics to predict the number of bugs in a file. The results of their predictions were significantly higher than real number of bugs. They found that the depth of inheritance is better than the compared metrics for the bug prediction, they only considered one single metric for the prediction. In~\cite{osman2018impact} authors  discussed the importance of choosing the correct metrics for the prediction of the number of bugs. They compared correlation based and wrapper based metrics for metric selection. They used different prediction models with multiple inputs for the considered models. They found that metric selection has an important effect on the prediction quality. The wrapper based metric selection found to be better than correlation based selection. However, they did not investigate correlation for different releases of the same project. This is an important issue because the similarity of the different releases of same project is expected to be higher and the metric selection based on that is expected to be more reliable. 

Instead of code metrics authors of \cite{fenton2008effectiveness} collected metrics depending on the judgement of the project managers of 31 different projects. These metrics includes information about design, documentation, scale of new code etc. Their method found more successful to predict the number of residual bugs of bigger projects compared to smaller projects. Although they could predict the number of bugs successfully the metrics they used is not easy to obtain and depends on personal comments of the project managers.

Instead of predicting the number of bugs some works focus on either a software is reliable or not. Reddivari and Raman \cite{reddivari2019software} applied different machine learning models for software quality prediction. They made binary decision about reliability of a system 'faulty' or 'not faulty' and classified its maintainability into three groups 'light', 'medium' and 'heavy' depending on code metrics. The Random Forrest algorithm found as a reliable classifier for both reliability and maintainability prediction. Visagan et al. \cite{visagan2019building} predicted either a software is reliable or not depending in its code metrics. They found that projects with high complexity have poor reliability. The binary prediction might be not considered as superficial for dependability analysis.

SRGM models are employed to predict the bug manifestation after the release date in \cite{vizarreta2017empirical}. It is been shown that these models can be used to predict the number of residual bugs some time after the release date of the new release. They also showed that this information can be used to make an interpretation on the reliability growth of the software.

\section{Proposed Framework}
\label{sec:fra}

Inevitably, any new release of a software project comes out with some bugs, that could not be found during its development and debugging period. These bugs, referred as residual bugs, have negative impact on the reliability of the new release. The goal of this framework is to provide a procedure to estimate the number of the bugs in the new release of a software project. The predicted value can be used to foretell the percentage of the bugs that are to be discovered and how reliable a new release is. Based on this estimation, developers/vendors can decide when the code is mature enough to be released and also users/customers can decide whether to  upgrade to the new release or keep the older version depending on their reliability requirements.

\begin{figure}[ht]
\centering
\includegraphics[scale=0.45]{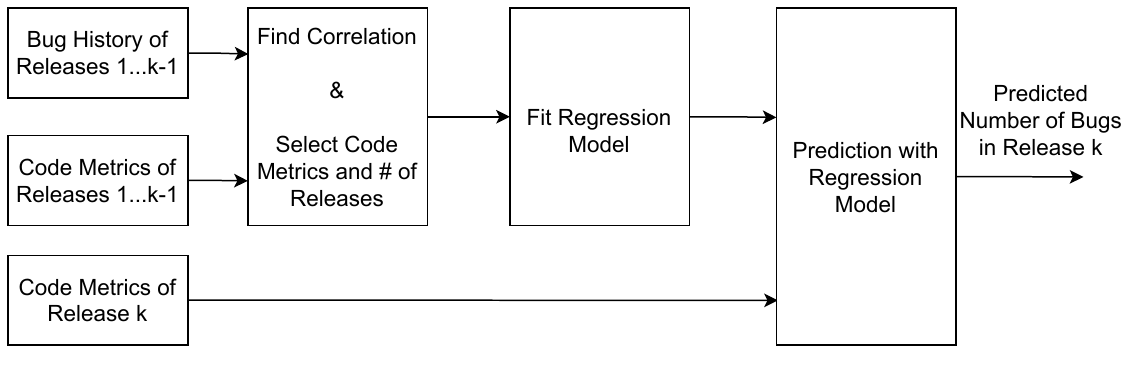}
\caption{Outline of the proposed framework to predict the number of the residual bugs of release $k$}
\label{fig:big_picture}
\end{figure}

Based on existing literature, newly added, removed features and modifications of the existing features have been identified as main contributors of residual bugs~\cite{zimmermann2009cross, nagappan2005use}. The proposed framework relies on the code metrics of the new and previous releases as well as on the bug history of this software project. Code metrics are measures that gives insights into the internal properties of the release (e.g., Number of lines, number of functions), as well as the development process of a particular software project (e.g., number of commits, number of developers). The proposed analysis relies on the correlation between these code metrics and the detected bugs of previous releases. The overview of the proposed framework is given in Figure~\ref{fig:big_picture}.

Most of the software projects follows a release cycle with dates that indicates: i) the start of the release ($t_s$), ii) start of the debugging period, so called code freeze date ($t_f$) and iii) release day ($t_r$). Between $t_s$ and $t_f$ new features are added to the software. At $t_f$, the debugging period starts and no more new features are allowed. Thus in this work the code from $t_f$ is considered as the code that contains all the bugs. 

Let us give an overview of the main steps to estimate the number of residual bugs on release $k$ given the bug history and the code metrics of previous releases: 
\begin{itemize}
    \item \textbf{\textit{Bug History:}} Many of current open-source software projects (e.g., ONAP, ONOS) use trackers like JIRA, Bugzilla and Mantis, which allow users and developers to report any bug they experience. These bug reports include detailed information such as creation date, bug description and current status of the bug (if it is still active, it has been solved, etc.). The \textit{bug history} in this framework represents the distribution of the bugs among the releases based on either the creation date or the affected release information in the bug report.

    \item \textbf{\textit{Code Metrics:}} The code metrics give information about the source-code features (e.g., number of lines of code) or development characteristics (e.g., number of commits) in a version control system. Most open-source software projects rely on version control systems like github and gitlab. These version control systems contain the history of the development process of the software. The code metrics of a particular software on a particular date can be found from the code on that date. Tools like SonarQube~\cite{sonar} and python libraries like pydriller~\cite{pydriller} support code metric calculations. 
    
    Some of the considered metrics like lines of code or number of files, provide fundamental information about the state of the code on a day. In this case the code from $t_f$ will be used to calculate the code metrics. On the other hand some code metrics like new lines of code aim to describe the development process, in other words only the changes made for the considered release. In order to calculate these metrics we take into account the changes made between $t_s$ and $t_f$. 

    \item \textbf{\textit{Bug History and Code Metrics Correlation:}} The proposed framework utilises the code metrics and bug history of releases $1...k-1$ to predict the number of bugs in the new release $k$ of a software project. This is done by finding a meaningful relation between the code metrics and bug history, and converting this relation into a regression model. This regression model together with the code metrics of the new release are used to find the number of bugs in a new release. For that purpose, the code metrics that correlate better along the bug history are preferred over the ones with lower correlation. Thus a valuable set of code metrics can be chosen with the help of correlation values.
    
    New projects might not have fair amount of previous releases to find a valuable correlation between bug history and code metrics. In case a project does not enough releases to introduce a meaningful correlation, similar software projects with same language, similar release cycle or code review structure can be used to fill this gap. Feeding the model with older releases might not increase the quality of the prediction, because the code structure, experience level of developers and writing style might change with time. Therefore a correlation analysis can be done before including all their releases.   

    
    \item \textbf{\textit{Fit Regression Model:}} Regression models are widely used for prediction and forecast~\cite{gupta2015developing}. A regression model defines a response variable $y$ related to one or list of covariates $x_1, x_2, ... , x_n $. The distribution of $y$ values is related to $x$ values and y value can be formulated as~\cite{fahrmeir2013regression}; 
    \begin{equation} \label{eq:reg_gen}
    y = f(x_1, x_2, ... , x_n) + \epsilon
    \end{equation}
    where $\epsilon$ corresponds to the error. From the given formula, the prediction value can be defined as $y = f(x_1, x_2, ... , x_n)$.  In the presented framework, regression model is based on the relation between the number of bugs in older releases $1 ... k-1$ (response variable ) and the code metrics of these releases $1 ... k-1$ (arguments). In this work fitting a regression model means finding a function that minimizes the error value $\epsilon$. Therefore the code metrics that offers high correlation with bug history are decent candidates to be used as arguments in the regression model.
    
    \item \textbf{\textit{Prediction with Regression Model:}} The fitted model then can be used to make a prediction of the target variable given the arguments of the new release. In other words, only the code metrics of the release $k$ are given to the regression model and regression model returns the predicted number of bugs of release $k$.

\end{itemize}

\section{Implementation}
\label{sec:imp}

In this section as an example one particular implementation of the proposed framework will be described. This implementation aims projects that use Jira for bug tracking and github for version control. The implementation of each step from Fig.~\ref{fig:big_picture} will be discussed in detail. The projects 'Open Network Automation Platform (ONAP)~\cite{onap} and Open Network Operating System (ONOS)~\cite{onos} are examined as example.

\subsection{Bug History}

Each software project using JIRA has a URL page. For example, the Jira page of the open-source project ONAP is ('https://jira.onap.org'), which  will be used as example. Since Jira is mostly used as project management tool, the type of tracked issues includes a wide variety of different types of issues (e.g., "Bug", "Milestone", "Improvement"). Some of them are also used for documentation of the software development steps (e.g., 'Milestone', 'Improvement' and 'Project Plan'). Since our purpose is to obtain the bug history, our implementation builds the \textit{bug history} by filtering all the issues of type "Bug". 

As an example, two of the bugs with their associated information have been shown in Table \ref{tab:jira}.

\begin{table}[ht]
\centering
\begin{tabular}[t]{ | l | c | c |}
\hline
Information & AAF-1192 & SO-3745\\
\hline
\hline
Sub-project Name & AAF & SO\\
\hline
Status & Closed & Open\\
\hline
Priority & High & Low \\
\hline
Affected Releases & Frankfurt, & Istanbul
\\& Guilin &  \\
\hline
First Affected Release & Frankfurt & Istanbul\\
\hline
Resolution & Done & Unresolved \\
\hline
Create Date & 2020-08-25 & 2021-08-24 \\
 & 14:27:21 & 14:30:09 \\
\hline
Create Date from Start & 1289 days & 1652 days\\
& 52041 seconds & 52209 seconds \\
\hline
Resolved Date & '2020-08-25 19:01:04' & None\\
\hline
Time to Solve (in hours) & 4.562 & None\\
\hline
Time to Solve (in days) & 0.19 & None\\
\hline
Months from Project Start & 42 & 54\\
\hline
Weeks from Project Start & 184 & 236\\
\hline
\end{tabular}
\caption{Examples of the collected information of two ONAP bugs from Jira}
\label{tab:jira}
\end{table}%

\begin{figure}[ht]
\centering
\includegraphics[width=8.8cm]{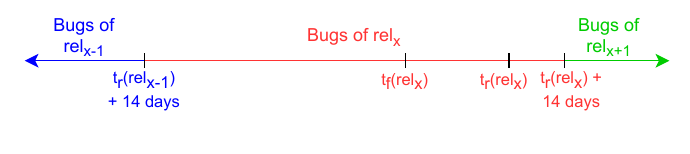}
\caption{{Important dates of a release and distribution of the bugs to the different releases}}
\label{fig:bug_rel}
\end{figure}

After filtering all the bugs, they are appointed to their corresponding release depending on their "First Affected Release" field information. Unfortunately, not all the bugs are properly defined and many bugs reported for ONOS do not have affected version in JIRA. In this case, those bugs are associated to the release based on their "Create Date" as most of them are found during the debugging period and coming weeks after its release date ($t_r$). Even if development of the new release $k+1$ is started, majority of the bugs after $t_r$ are related to newly released release $k$. Thus if the creation date of the bug is 14 days after $t_r$ of release $k$, the first affected version as considered as release $k$. The dates and ditribution of the unlabeled bugs to releases are visualized in Fig. \ref{fig:bug_rel}.

\subsection{Code Metrics}

The proposed implementation calculates the code metrics by analyzing the  code. Therefore the code has to be downloaded from Github before code metric extraction. 

Also some metrics are calculated as sum of the other metrics like number of new and modified LoC.

\subsubsection{Downloading the repositories from Github}

Two dates $t_s$ and $t_f$ are considered to extract the code metrics. It is assumed that the bugs appear because of the newly added features and changes made between $t_s$ and $t_f$. Thus all the bugs of the release is assumed to introduced before $t_f$ in other words the code from $t_f$ is expected to have all the bugs of the considered release.

Some of the code metrics that considered are measures of the state of the code like number of functions, total complexity of the project and number of thousand lines of code (kLoC). The code state from $t_f$ is enough to get those information. On the other hand having the code status from one day is not enough to find some of the considered metrics which measures the changes made in the code. New kLoC, number of new functions are some examples to these metrics, that cannot be found from one state of the code. Since the changes related to the release are made after $t_s$, the code state from this date is considered as the old state of the code and state from $t_f$ is considered as the new version. Therefore the code states from last github commit before $t_s$ and the last commit on $t_f$ are downloaded. the commits added after $t_f$ are considered as part of debugging process. The $t_s$ and $t_f$ dates should be found in the website or documentation of the considered projects and given by user.

Some of the code metrics like number of commits or number of contributors cannot be found in source-code instead they should be gathered from github. The number of commits per release is also extracted during the code download process. In this work the metric 'commit' represents the number of commits between $t_s$ and $t_f$ like other metrics that measures change.



\subsubsection{Extracting the code metrics}

After downloading the source-code of different releases from the github, they are analyzed to extract the code metrics. In this work the code metrics will be divided into two categories. The terms complexity metrics and code size metrics will be used to differentiate these categories. The complexity metrics are calculated from each function separately and they are metrics related to their cyclomatic complexity like number of functions with complexity value higher than 15. The code size metrics are the metrics which defines the length of code like number of thousand lines of code or the changes made in the code like thousands of new lines of code and modified lines of code. Two different tools are used for the calculation of the metrics of these categories. The code size metrics are found with Count Lines of Code (CLOC) \cite{cloc}. It gives the information about number of lines of code, number of files and also calculates the differences between two folders in this manner. Lizard \cite{lizard}, which makes function level analysis, is used for complexity metrics. It calculates the complexity of the functions and also used to calculate the number of functions and number of new functions.

\paragraph{Code Size Metrics}

The code size metrics describe the metrics which relate to the length of the code and quantity of the changes made in the code. We used CLOC tool \cite{cloc} to calculate the code size metrics, because it calculates not only the lines of code but also the differences between the code in two different directories. The CLOC tool goes through all files and returns LoC, new LoC, modified LoC, number of files, number of new files $etc.$. This part takes a directory as input and returns the related metrics as output. 

The projects that are analyzed in this work are Java based projects with around 70\% of all the LoC is written in Java. Therefore the code size metrics are extracted also specific to files written in Java. The languages like 'yaml', or 'xml' are excluded, because they mostly do not contain logical operations. Since they are mostly do not include complex operations, it has been assumed that they are mostly not the reason behind the bugs and even some of them like might be created automatically.

\paragraph{Complexity Metrics}

Complexity related information are collected with the help of the 'Lizard' Python library \cite{lizard}. The input of our implementation is the directory of the considered project and as output it gives the complexity metrics for this directory. In this work cyclomatic complexity is used to evaluate the code complexity, which is found useful by many research and widely used~\cite{ebert2016cyclomatic}. It represents the count of linearly independent paths in a piece of code. 

The counts of the functions with complexity values higher than specific thresholds (10, 15 and 20) are one of the extracted complexity related metrics. Additionally, the summation of the complexity numbers of all functions is considered. The main purpose of this tool is calculating the complexity of functions meanwhile it goes through all of the functions in the analyzed file. Thanks to this strategy the number of functions is also calculated.

In order to get complexity metrics for new and modified functions, the 'filecmp' Python library is used. First newly added and changed files in the project are found, then all the functions in the mew files are added to the count of new and modified functions. All functions in modified files are compared and if they are new or modified, they will also added to the count of new and modified functions.


\subsection{Bug History and Code Metrics Correlation}

The code metric extraction returns 43 different code metrics. Since the number of metrics is high, some of them should be preferred over the others to predict the number of bugs in the new release. A meaningful relation between code metrics and bug history has been searched to choose the code metrics for prediction. Pearson correlation coefficient (PCC) is a measure for linear correlation to evaluate the correlation between two sets of data~\cite{eid2013linear}. In order to evaluate the correlation between bug history and code metrics PCC is used. PCC gets a value between $-1$ and $+1$, where $0$ means no correlation between two sets of value. Positive PCC value indicates positive correlation and negative PCC value means that correlation is negative. Higher absolute value of PCC indicates stronger correlation, where $-1$ and $+1$ point to perfect correlation. If the absolute PCC value is higher than 0.7, the correlation is considered as "high" correlation, whereas a value between 0.4 and 0.7 indicates "significant" correlation~\cite{yu2017research}. The correlation is weak if the PCC is below 0.4. The metrics with higher absolute PCC are expected to be more valuable for prediction. In this work 'pandas' library \cite{pandas} is used to calculate the correlation between bug history and code metrics. 

\subsection{Fit Regression  Model}

The experience from older releases is used to calculate the number of the bugs in the new release. The code metrics that has higher correlation with number of bugs has been chosen to fit the regression model. Since PCC gives information about linear correlation, linear regression model is preferred. The formula of the linear regression model is given in the following expression~\cite{fahrmeir2013regression}:

\begin{equation} \label{eq:multi_reg}
y_i = \beta_0 + \beta_1x_{i1} ... + \beta_nx_{in} + \epsilon_i,\hspace{0.4cm}i = 1, ..., k-1
\end{equation}

where $\beta_0$ is the intercept, $\beta_1, \beta_2, ... , \beta_n$ are the slopes of the line and $\epsilon_i$ is the error value. In this work $k-1$ represents the number of older releases used to fit the regression model, $y_i$ is the number of bugs in release i and $x_{in}$ represent the $n$th code metric value for release $i$. Fitting the regression model aims at finding the optimal $\beta$ values that minimize the error $\epsilon$ for all the older releases. In this work 'scikit-learn' library~\cite{scikit} is used to fit the regression model. Code metrics of the older releases and number of bugs of the same old releases are given to the regression model to fit.

\subsection{Prediction with Regression Model}

After the model is fitted the code metrics of the new release is used to estimate the number of bugs in the new release. The calculated $\beta$ values are express the fitted regression model. The code metrics values of the new release k ($x_{k1}, ... , x_{kn}$) are given to the fitted model to predict the number of bugs in release k $(y_k)$. Since the $\beta$ values are chosen to minimize the error $\epsilon_i$ for the older releases, $\overline{y_k} $ value is expected to be close to the real $y_k$ value. 

\section{Results}
\label{sec:res}

The implementation described in Section~\ref{sec:imp} is applied to two different open source software projects ONOS and ONAP. In this section the results for each step of the framework will be given.


\subsection{Bug History}

Let us start with the ONAP project, whose first release Amsterdam was released on 16 November 2017. Since then, 10 more releases have been delivered. An ONAP release from its start $t_s$ to its release $t_r$ takes approximately 6 months. From its start to 17 October 2022, 11.666 bugs have been reported in ONAP Jira. These bugs have been assigned to their corresponding release based on the bug's affected release label in Jira.However, around $9.5\%$ of these bugs could not be assigned as they do not have an information about the first affected release. Those bugs are assigned to the releases based on the issue creation date. The collected information for two example bugs are given in Table~\ref{tab:jira}.


\begin{figure}[ht]
\centering
\includegraphics[width=8.2cm]{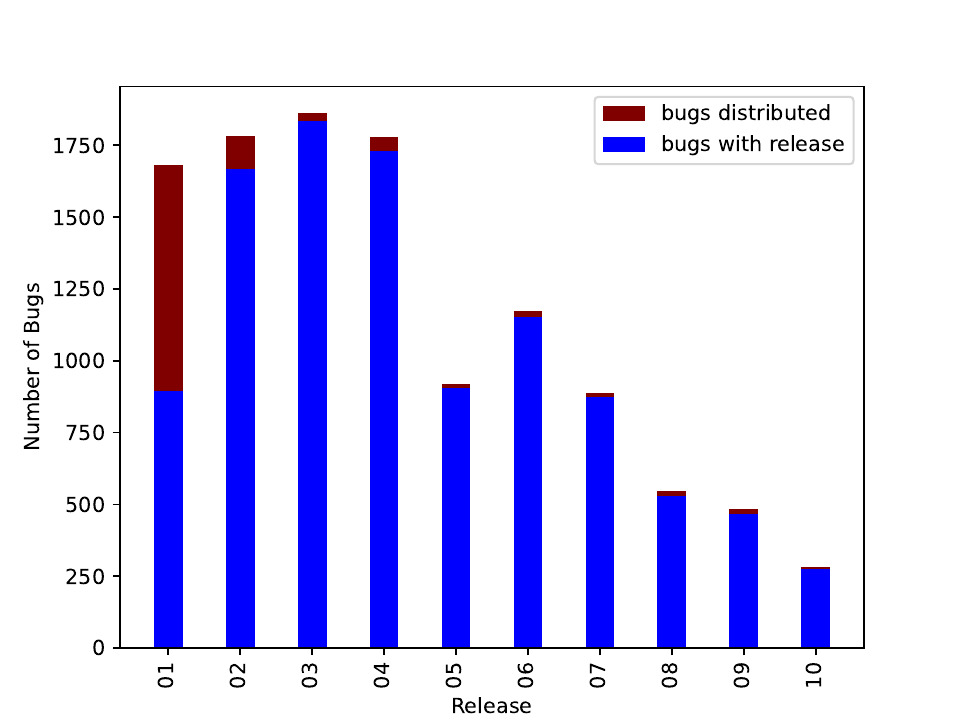}
\caption{{Bug count per ONAP release}}
\label{fig:bug_hist_onap}
\end{figure}

As of 17 October 2022, ONOS project has 24 releases. An ONOS release from  $t_s$ to $t_r$ takes approximately 4 to 5 months. Similarly to ONAP, the bugs have been mapped to their corresponding release according to the collected bug information. In ONOS, $45\%$ of all 2301 bugs (17 October 2022) do not have an affected release, and hence, those bugs are associated to the release based on their create dates. ONOS also has long time support (LTS) releases, whose bugs can be found for longer time compared to normal releases. It make the distribution process of the unlabeled bugs less convenient compared to ONAP. 

\begin{figure}[ht]
\centering
\includegraphics[width=8.2cm]{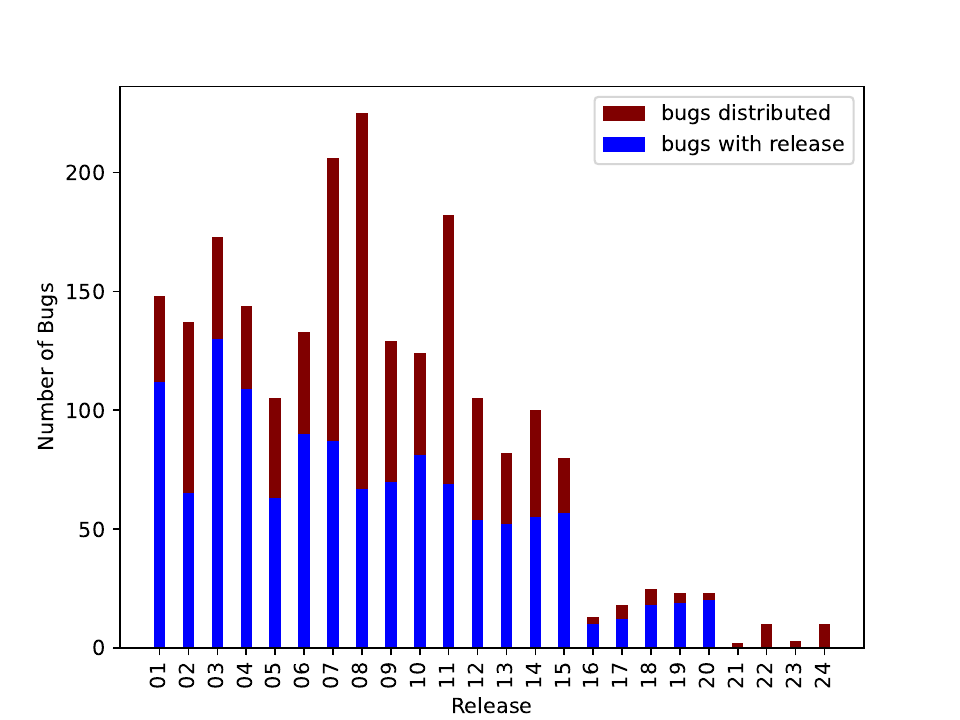}
\caption{{Bug count per ONOS release}}
\label{fig:bug_hist_onos}
\end{figure}

The number of bugs for the different releases of the ONAP and the ONOS projects are depicted in Figures~\ref{fig:bug_hist_onap} and Figure~\ref{fig:bug_hist_onos} respectively, where the blue bars correspond to the bugs with release information and the brown bars correspond to the associated bugs based on the bug creation date. It can be observed that for both projects, the number of bugs per release are higher for the first releases. However, there is a difference in the number and distribution of bugs without the release label. In ONAP, most of those bugs are associated to the first release, whereas for ONOS, those bugs are distributed along most of the releases. Also first ONAP releases contain the largest number of bugs without the affected release label but this trend is not observed in ONOS.


\subsection{Code Metrics}

6 examples of the 43 code metrics are chosen as example. Those are the examples of development effort (commits), code metrics based on project size (kLoC), size of changes made in the code (new, removed kLoC), changes made in Java code (new kLoC Java) and addition of the complex functions (new complex Java functions). 

The number of bugs and 6 code metrics of the 6 ONAP releases are given in Table~\ref{tab:metrics_onap}. Release 1 is chosen because it is the first release, release 3 is the most buggy release and 10 is the last and least buggy release. Other three release are intermediate points between those releases. Here it can be seen that the release 7 has the highest number of kLoC but it does not have the highest or lowest number of bugs. On the other hand, release 3 has the most number of commits and removed kLoC also the highest number of bugs. The release with the lowest number of bugs (release 10) is the one with the lowest number of commits. It can be said that there is a correlation between the code metrics and number of bugs. From this trend it can be said that if the number of commits and removed LoC is high, number of bugs in this release is expected to be high.

\begin{table}[ht]
	\centering
	\begin{tabular}[t]{ | l || c | c | c | c | c | c |}
		\hline
		Release & 01 & 03 & 05 & 07 & 09 & 10\\
		\hline
		\hline
		Bugs & 1.681 & 1.862 & 920 & 887 & 484 & 282\\
		\hline
		Commits & 7.474 & 10.504 & 7.138 & 4.186 & 2.675 & 2.364\\
		\hline
		kLoC & 8.833 & 7.961 & 9.602 & 10.347 & 9.624 & 8.850\\
		\hline
		new kLoC & 5.599 & 2.590 & 1.142 & 1.192 & 374 & 395\\
		\hline
		removed & \multirow{2}{*}{2.027} & \multirow{2}{*}{2.399} & \multirow{2}{*}{579} & \multirow{2}{*}{397} & \multirow{2}{*}{165} & \multirow{2}{*}{383}\\
		kLoC &  &  &  &  &  & \\
		\hline
		new kLoC & \multirow{2}{*}{1.087} & \multirow{2}{*}{545} & \multirow{2}{*}{142} & \multirow{2}{*}{190} & \multirow{2}{*}{107} & \multirow{2}{*}{48}\\
		Java &  &  &  &  &  & \\
		\hline
		new complex & \multirow{2}{*}{897} & \multirow{2}{*}{179} & \multirow{2}{*}{52} & \multirow{2}{*}{32} & \multirow{2}{*}{30} & \multirow{2}{*}{3}\\
		Java function &  &  &  &  &  & \\
		\hline
	\end{tabular}
	\caption{6 Example Code Metrics for 6 ONAP releases}
	\label{tab:metrics_onap}
\end{table}%

Same example code metrics for the 6 releases of the ONOS project are given in Table~\ref{tab:metrics_onos}. Also first, last, most buggy (release 8) and least buggy (release 21) releases are chosen for ONOS. Other two releases are the releases between these releases. In ONOS also the release with the lowest number of bugs (release 2) has the lowest number of commits and removed kLoC. The release with highest number of bugs (release 8) has the highest number of bugs but it does not has the highest number of commits. This shows that code metrics can be used to predict the number of bugs in a release in ONOS project.

The kLoC value gives broad information about the size of the project and number of commits and new kLoC reflects the development effort for the related release. From Tables~\ref{tab:metrics_onos} and~\ref{tab:metrics_onap}, it can be said that the ONAP project is much bigger and more development efforts have been made for the given releases.  In connection with the development efforts the number of bugs of the ONAP releases are also higher than the ONOS releases.

\begin{table}[ht]
	\centering
	\begin{tabular}[t]{ | l || c | c | c | c | c | c |}
		\hline
		Release & 01 & 04 & 08 & 15 & 21 & 24\\
		\hline
		\hline
		Bugs & 148 & 105 & 225 & 80 & 2 & 10\\
		\hline
		Commits & 383 & 668 & 607 & 598 & 98 & 74\\
		\hline
		kLoC & 124 & 679 & 1.105 & 1.497 & 1.615 & 1.522\\
		\hline
		new kLoC & 43 & 483 & 73 & 156 & 6 & 22\\
		\hline
		removed & \multirow{2}{*}{2} & \multirow{2}{*}{38} & \multirow{2}{*}{56} & \multirow{2}{*}{49} & \multirow{2}{*}{2} & \multirow{2}{*}{5}\\
		kLoC &  &  &  &  &  & \\
		\hline
		new kLoC & \multirow{2}{*}{11} & \multirow{2}{*}{76} & \multirow{2}{*}{50} & \multirow{2}{*}{54} & \multirow{2}{*}{5} & \multirow{2}{*}{6}\\
		Java &  &  &  &  &  & \\
		\hline
		new complex & \multirow{2}{*}{1} & \multirow{2}{*}{15} & \multirow{2}{*}{7} & \multirow{2}{*}{22} & \multirow{2}{*}{0} & \multirow{2}{*}{1}\\
		Java function &  &  &  &  &  & \\
		
		\hline
	\end{tabular}
	\caption{6 Example Code Metrics for 6 ONOS releases}
	\label{tab:metrics_onos}
\end{table}%

\begin{figure}[!h]
  \centering
  \subfloat[ONAP]{\includegraphics[clip, width=.45\textwidth, trim= 0cm 0cm 0cm 0cm]{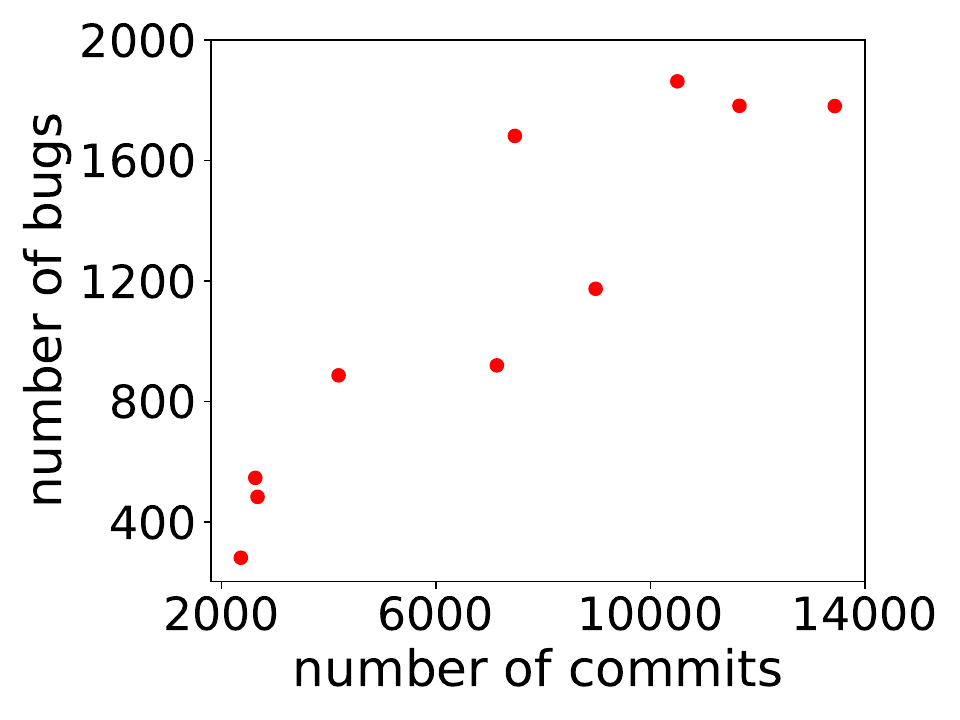}\label{fig:onap_commit_bug}}\\
  \subfloat[ONOS]{\includegraphics[clip, width=.45\textwidth, trim= 0cm 0cm 0cm 0cm]{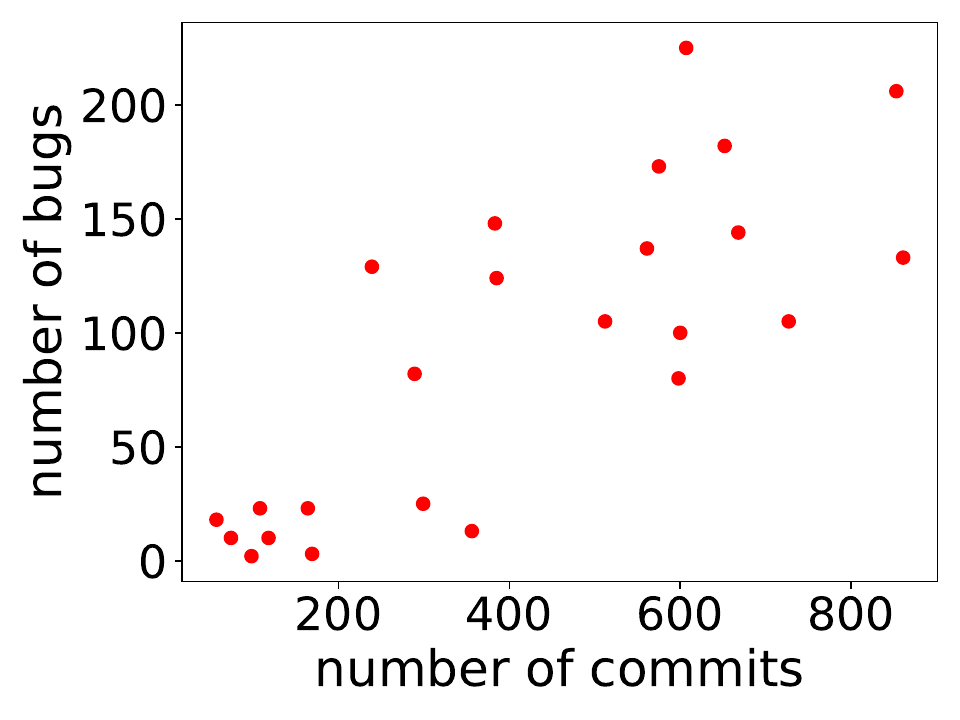}\label{fig:onos_commit_bug}}\\
  \caption{Change of number of bugs with number of commits}\label{fig:commit_bug}
\end{figure}

To understand the connection between code metrics and bug history better, the change of bug history and commits that corresponds to the development efforts is shown in Figure~\ref{fig:commit_bug}. It can be observed, that ONAP is closer to a linear dependence than ONOS. 

\subsection{Bug History and Code Metrics Correlation}
\begin{figure}[ht]
\centering
\includegraphics[width=8cm]{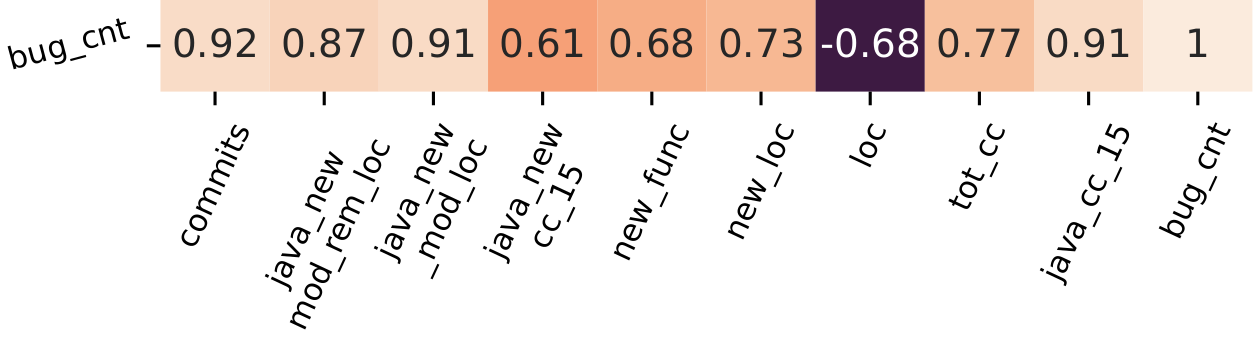}
\caption{{Correlation between number of bugs and code metrics for ONAP}}
\label{fig:cor_onap}
\end{figure}

\begin{figure}[ht]
\centering
\includegraphics[width=8cm]{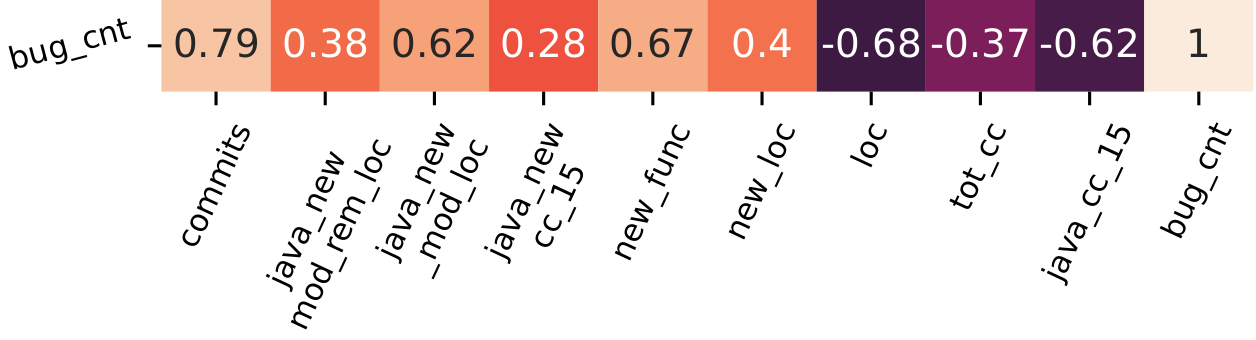}
\caption{{Correlation between number of bugs and code metrics for ONOS}}
\label{fig:cor_onos}
\end{figure}

The correlation between the bug history and the code metrics have been performed for ONAP and ONOS as shown in Figures~\ref{fig:cor_onap} and ~\ref{fig:cor_onos} respectively. 

As observed in Figure~\ref{fig:commit_bug}, the expected number of bugs increases proportionally with number of commits. In order to evaluate the correlation between the number of bugs and each code metric, the Pearson correlation coefficient (PCC) is used.

Let us focus on the correlation of the code metrics with the bugs for ONAP, which are shown in the last row of the Figure~\ref{fig:cor_onap}. Six out of nine metrics shown "high" correlation, and in particular metrics like \textit{commits, sum of new, modified LoC} and \textit{number of complex function} have a correlation coefficient higher than 0.9. Those "high" correlation values motivate the use of code metrics to predict the number of bugs in a new ONAP release.

However, the correlation of ONOS is different as shown in Figure~\ref{fig:cor_onos}. In this case, the PCCs for code metrics and the number of bug are lower than the ONAP project. Only the metric \textit{commits} has "high" correlation and three other metrics have "significant" correlation. Since the correlation values are lower than ONAP project, the prediction accuracy is expected to be also low.

\subsection{Regression Model}
\label{subsec:reg}

Linear regression model, whose formula is given in Formula~\ref{eq:multi_reg}, is used to predict the number of bugs. Fitting the regression model aims at finding the optimal $\beta$ values that minimize the $\epsilon$ values.

\begin{figure}[ht]
\centering
\includegraphics[width=8cm]{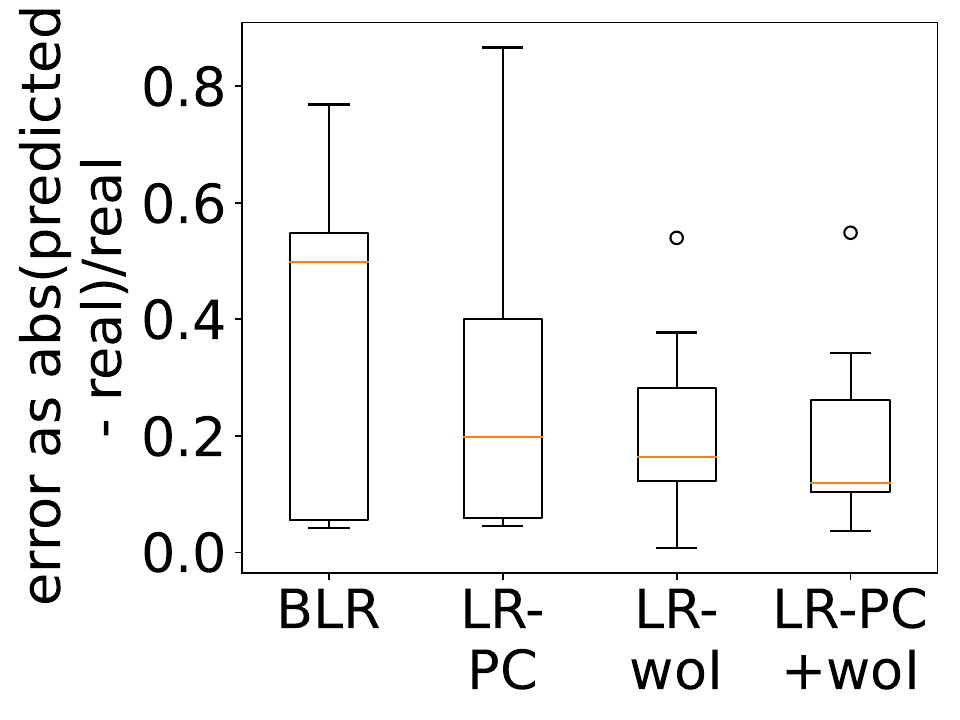}
\caption{{Prediction error of different adjustments of the linear regression model for ONAP}}
\label{fig:regr_onap}
\end{figure}

\begin{table}[ht]
	\centering
	\begin{tabular}[t]{ | l | l || c | c | c | c |}
	\hline
		 				&	& median & mean & max & min\\
	\hline
	\hline
	\multirow{2}{*}{BLR} & ONAP & 0.499  &  0.37  &  0.769  &  0.042\\
		\cline{2-6}
		& ONOS & 0.737  &  2.228  &  18.63  &  \textbf{0.038}\\
	\hline
		\multirow{2}{*}{LR-PC} & ONAP & 0.199  &  0.293  &  0.866  &  0.045\\
		\cline{2-6}
		& ONOS &  0.49  &  2.357  &  17.143  &  0.062\\
	\hline
		\multirow{2}{*}{LR-woI} & ONAP & 0.164  &  0.206  &  \textbf{0.54}  &  \textbf{0.008}\\
		\cline{2-6}
		& ONOS & 0.837  &  2.942  &  21.768  &  0.049\\
	\hline
		\multirow{2}{*}{LR-PC+woI} & ONAP & \textbf{0.12}  &  \textbf{0.195}  &  0.549  &  0.036\\
		\cline{2-6}
		& ONOS & \textbf{0.486}  &  \textbf{1.634}  &  \textbf{11.476}  &  \textbf{0.038}\\
	\hline
	\end{tabular}
	\caption{Median, mean, maximum and minimum prediction error values of different adjustment of the linear regression model for ONAP and ONOS}
	\label{tab:regr_onap_onos}
\end{table}%

The number of commits, number of new LoC, sum of new and modified Java LoC, number of new functions and number of complex Java functions values are used to feed the regression model as they offer higher correlation values and they give information about different aspects like development effort, complexity and Java based changes. The prediction values are evaluated with an error value which calculated with given formula:

$$e = \frac{|E[\#Bugs] - \#Bugs|}{\#Bugs} $$

where $e$ represents the error value, $E[\#Bugs]$ is the predicted number of bugs and $\#Bugs$ is the real number of bugs.

Figure~\ref{fig:regr_onap} shows the distribution of the error value of the linear regression for different adjustments. In the first case referred as base linear regression model (BLR), all the previous releases are considered to feed the model (e.g., to predict the bugs of the fifth release, all releases up to release four are used to fit the linear regression).

Since all the chosen code metrics have positive correlation with bug history of ONAP, the second case referred as "linear regression with positive coefficients" (LR-PC) considered only positive slope $\beta$ values. Also a linear regression model without intercept value (LR-woI) $\beta_0 = 0$ (third case) is trained because the changes made in the code is considered as the reason of the emerging bugs and without any change no bug is expected. The drop in the median value of error can be seen in the Figure~\ref{fig:regr_onap} 

In order to get better insight of the figure, Table~\ref{tab:regr_onap_onos} summarizes the mean, median, maximum and minimum error values for each case. Both median and mean values are lower for LR-PC and LR-woI compared to BLR model. A "linear regression model without intercept and positive slope (LR-PC+woI) offers even lower error values. The median value drops from 0.499 to 0.12 which shows that the prediction can become closer to the real value with finding the correct structure of the linear regression model. These results show that the LR-PC+woI can be used to predict the number of bugs in the new release of the ONAP project.

\begin{figure}[ht]
\centering
\includegraphics[width=8cm]{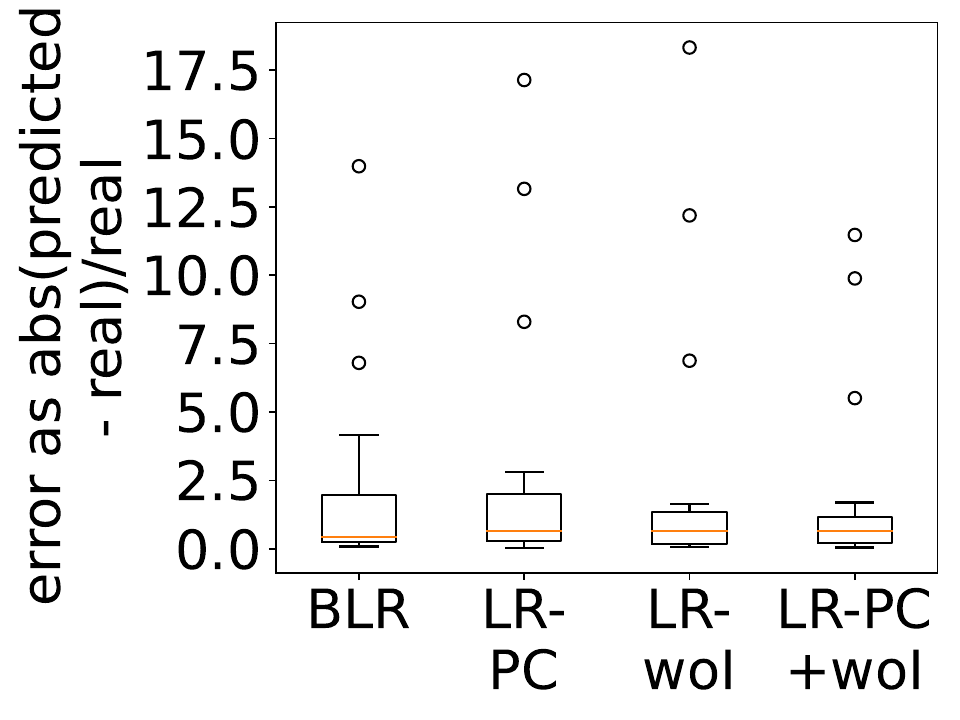}
\caption{{Prediction error of different adjustments of the linear regression model for ONOS}}
\label{fig:regr_onos}
\end{figure}

Similarly to ONAP, all code metrics except for the number of complex Java functions, have been used to fit the regression model to ONOS releases. The reason to exclude the number of complex Java functions is due to its negative correlation with the number of bugs and using a positive slope is only meaningful if all the correlations are positive. Instead sum of new, modified and removed LoC of Java is used. 

The box plot of the error value for the predicted number of bugs with different compositions for ONOS is given in Figure~\ref{fig:regr_onos}. It can be observed that there are more outliers than in ONAP. The not labeled data and irregular release dates might cause the very high errors.

The mean and median values of the prediction error are given in Table~\ref{tab:regr_onap_onos}. The lowest error value is reached with using positive slopes and no intercept for both projects. On the other hand using positive slopes alone or no intercept does not give lower error value. The higher values show that the lower correlation value effects the prediction accuracy negatively. 

\subsection{Cross-Project Prediction}

Sometimes a project does not have enough previous releases to fit a regression model and achieve a realistic prediction of the number of bugs. As an example, the ONAP project started in November 2017, 3 years later than ONOS project. Can this model be used to predict the number of bugs in a new project with the help of older project? In order to answer this question, a regression model is fitted to all the ONAP and ONOS releases which were released before the code freeze date of the predicted release of the ONAP project. As example, the first 11 ONOS releases are used to predict the first ONAP release and the regression model is fitted to fitst ONAP release and first 13 releases ONOS releases that released before second release of ONAP to predict the second ONAP release. The results from the Table~\ref{tab:regr_onap_from_onos} shows that the predicted values can give a rough idea about the number of bugs in the release. TheSince the linear regression model with positive slope and without intercept was better for in-project prediction the results of these model will be discussed. The error value is 0.081 for the first release, 0.481 for the second and 0.11 for the third release. The cross-project strategy can be used for the prediction of the number of bugs in the first releases of a new software.


\begin{table}[ht]
	\centering
	\begin{tabular}[t]{ | l || c | c |}
		\hline
		 & ONAP + ONOS & ONOS\\
		\hline
		\hline
		Release 1 & 0.081  &  - \\
		\hline
		Release 2 & 0.481  &  0.549\\
		\hline
		Release 3 & 0.11  &  1.17\\
		\hline
		Release 3 & 0.113  &  0.104\\
		\hline
	\end{tabular}
	\caption{Prediction of first ONAP releases with help of ONOS and previous ONAP releases alone}
	\label{tab:regr_onap_from_onos}
\end{table}%

\subsection{Prediction Accuracy with Different Number of Considered Releases}

From release to release, the experience level of the contributors, the release procedures or priorities of the teams can change, which may impact the number of bugs. Over time, these aspects may affect the similarity between the releases and hence, the objective of this study is to evaluate how many of the previous releases should be considered to achieve the best estimation of the number of bugs of a new release. 
In order to perform this evaluation, the first task is to evaluate the correlation of the different metrics for the different number of releases. Based on this correlation, the change in the correlation value for different number of releases will be given.

\begin{table}[ht]
\centering
\begin{tabular}[t]{ | c || c | c | c | c | c |}
\hline
 number of &   & median &  & mean \\
 previous &  median & new + modif &  mean & new + modif \\
 releases & commits & LoC Java & commits & LoC Java \\
\hline
\hline
2  &  0.895  &  0.986  &  0.688  &  0.659 \\
\hline
3  &  0.92  &  0.985  &  0.86  &  0.739 \\
\hline
4  &  0.893  &  0.979  &  0.87  &  0.947 \\
\hline
5  &  0.914  &  0.977  &  0.873  &  0.949 \\
\hline
6  &  0.936  &  0.979  &  0.907  &  0.952 \\
\hline
7  &  0.946  &  0.974  &  0.921  &  0.949 \\
\hline
8  &  0.927  &  0.938  &  0.927  &  0.938 \\
\hline
9  &  0.916  &  0.912  &  0.916  &  0.912 \\
\hline
\end{tabular}
\caption{Median and mean PCC for bugs history and code metrics for different number of considered releases for ONAP}
\label{tab:cor_rel_onap}
\end{table}%

As first task, we evaluate the PCC for a different number of releases. The median and mean PCC for bug history and two code metrics, which offer higher correlation value, commits and new and modified number of Java code is given are shown in Table~\ref{tab:cor_rel_onap}. The number of previous releases is the number of releases used to get the correlation coefficient. Let us give as an example the case of considering 2 previous releases. In this case, for release $x$, the code metrics of releases $x-1,x-2$ are considered.  
The median and mean values are calculated from all the correlation coefficients values that are gathered using the same number of releases.
From this Table~\ref{tab:cor_rel_onap}, it can be seen that the median values are high for lower number of releases, it is because those releases are coming from a shorter time period and similar to each other. On the other hand, the mean value is lower because a smaller shift in one release causes a lower correlation value. This is because of the low mean value using 2 previous releases might give lower prediction accuracy. After 3 previous releases, both mean correlation coefficients becomes higher than 0.7 which is the limit for high correlation. Although the mean values are comparably higher for more releases the median value does not follow the same trend. The median value for the best correlated metrics drops from 0.979 to 0.974. Thus using more than 3 and less than 7 previous ONAP releases seems reasonable from the correlation values. 

\begin{figure}[ht]
\centering
\includegraphics[width=9.5cm]{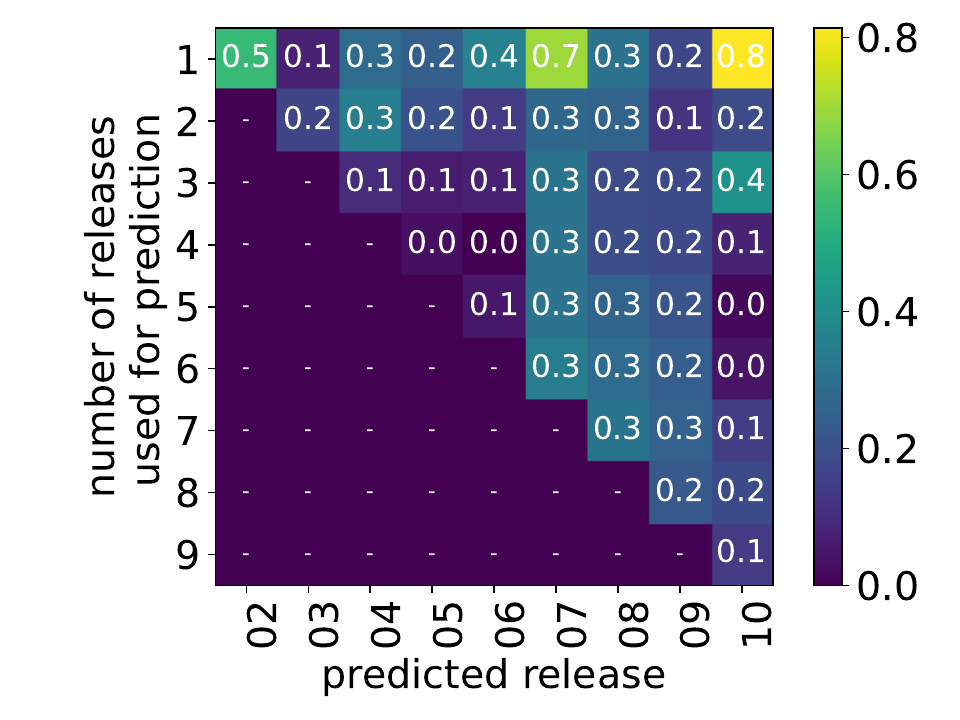}
\caption{{Prediction error of different number of releases used for ONAP}}
\label{fig:pred_rel_onap}
\end{figure}

The error value for prediction of each ONAP releases with different number of previous releases used to fit the regression model is given in Figure~\ref{fig:pred_rel_onap}. Linear regression model with positive slope values and without intercept is used for prediction. There are more results for latest releases because for those releases more  previous releases can be given to the model. After using more than 3 previous releases none of the prediction errors reach to 0.4, on the other hand after 7 previous releases all the error values are higher than 0.1. It can be said that neither using all the releases nor using only the last one or two releases gives the optimal prediction performance. 

\begin{table}[ht]
\centering
\begin{tabular}[t]{ | c || c | c | c | c | c | c |}
\hline
 number of &  &  &  &  & median &  mean\\
 considered & median & mean & max & min & last 4 & last 4\\
  release  &  &  &  &  & releases & releases \\
\hline
\hline
1 & 0.302  &  0.389  &  0.814  &  0.075 & 0.498  &  0.497 \\
\hline
2 & 0.195  &  0.214  &  0.349  &  0.126 & 0.222  &  0.211 \\
\hline
3 & 0.183  &  0.191  &  0.416  &  0.061 & 0.248  &  0.274 \\
\hline
4 & 0.129  &  0.133  &  0.309  &  0.001 & 0.188  &  0.19 \\
\hline
5 & 0.214  &  0.171  &  0.309  &  0.016 & 0.238  &  0.2 \\
\hline
6 & 0.264  &  0.228  &  0.342  &  0.044 & 0.264  &  0.228 \\
\hline
7 & 0.255  &  0.239  &  0.316  &  0.147 & 0.255  &  0.239 \\
\hline
8 & 0.204  &  0.204  &  0.228  &  0.181 & 0.204  &  0.204 \\
\hline
9 & 0.141  &  0.141  &  0.141  &  0.141 & 0.141  &  0.141 \\
\hline
\end{tabular}
\caption{Median, mean, maximum and minimum prediction error values with different number of releases used to regression fit for ONAP}
\label{tab:pred_err_rel_onap}
\end{table}%

The mean and median error values for regression model, which feed with different number of releases is given in Table~\ref{tab:pred_err_rel_onap}. In order to make a fair comparison prediction error for the last 4 releases is also given because the error values is comparably higher than 4th, 5th and 6th releases as it can be seen at the Figure~\ref{fig:pred_rel_onap}. The best value is 0.141 of the prediction with 9 previous releases but it only has one sample and the error for this sample is not lower than prediction with 4, 5 or 6 previous releases which can be seen in the last column of the Figure~\ref{fig:pred_rel_onap}. After 9 previous releases the lowest number is reached with using 4 previous releases which followed by 5 previous releases. If we compare this results with the results in Subsection~\ref{subsec:reg}, we can see that the mean value of the 4 previous releases is lightly higher but the mean value is comparably lower.

\section{Conclusion}
\label{sec:con}

This paper proposes a framework to predict the number of bugs in a new software release. First each step of the proposed framework has been described. Then an exemplary implementation of this framework has been discussed. Two open-source projects ONOS and ONAP were analyzed with the presented implementation. The prediction has been improved with different methods and cross-project prediction is also tested to predict a new software with low number of old releases. 

The results shows that the number of bugs can be predicted at the end of the process of the software. The trustworthiness of this prediction depends on the project's properties. When ONOS and ONAP projects are compared, it can be seen that the correlation between code metrics and bug history of ONAP project is relatively higher, which leads to better prediction accuracy. One reason for that can be the number of bugs, which does not have any effected release information in issue tracker, is higher in ONOS. Other possible reasons are sliding release times of ONOS project and different type of releases like normal and long time support. The results also show that the bug history of a similar project can give a rough prediction on the number of releases of the first releases of the software. The optimal number of releases that should be used for a better prediction is also investigated. 

As future work this framework can be taken forward to predict the number of bugs in the su-bmodules of a software project. Also it can be tested with different machine learning models. The distribution of the unlabeled bugs can be improved with further investigation of commits that solves the bugs. Bug reports also includes information about the priority of the bugs, the framework can be tested with the bugs of different priority. A software reliability growth model can be feed with the predicted number of bugs and the expected reliability of the software on release date can be foreseen.

\section*{Acknowledgment}

This work has received funding from the "DFG" under grant numbers MA 6529/4-1 and KE 1863/10-1.

\end{document}